# THE GEOMETRIC GRAVITATIONAL INTERNAL PROBLEM.

### Gustavo González-Martín

Departamento de Física, Universidad Simón Bolívar,

Apartado 89000, Caracas 1080-A, Venezuela.

Web page URL http:\\prof.usb.ve\ggonzalm\

In a geometric unified theory there is an energy momentum equation, apart from the field equations and equations of motion. The general relativity Einstein equations with cosmological constant follow from this energy momentum equation for empty space. For non empty space we obtain a generalized Einstein equation, relating the Einstein tensor to a geometric stress energy tensor. The matching exterior solution is in agreement with the standard relativity tests. Furthermore, there is a Newtonian limit where we obtain the corresponding Poisson's equation.

# Introduction.

The standard approach to internal gravitational solutions rests on the construction of an energy momentum or stress energy tensor *T* determined by the classical equations of macroscopic matter and fields. In general this matter fields are described by the classical fluid equations. In particular, in the Einstein Maxwell theory, this tensor has a field contribution that may be taken as the electromagnetic energy momentum tensor $T_E$. Nevertheless, the quantum aspects of matter are not explicit in this construction. On the other hand, it may claimed that these theories are not truly geometrically unified theories. Einstein [1] himself was unsatisfied by the non geometrical character of *T* and spent his later years looking for a satisfactory unified theory. A geometric unified theory that incorporates the quantum aspects of matter has been presented [2,3].

Associated to this unified theory there is a physical geometry that provides three equations for the objects of the theory. The first or field equation is a differential equation for the generalized connection with a current matter source. The second determines the motion of source objects in terms of the covariant derivative. The third

$$\mathrm{tr}\left[2\Omega_{\hat\rho\nu}\Omega^{\mu\nu} - \tfrac{1}{2}u^\mu_{\hat\rho}\Omega^{\kappa\lambda}\Omega_{\kappa\lambda}\right] = 2k\,\mathrm{tr}\left[e^{-1}\iota\circ u^\mu\nabla_{\hat\rho}e\right] \qquad (1.1)$$

has the structure of a sum of terms, each of the form corresponding to the electromagnetic stress-energy tensor. A tensor with the structure of stress-energy may be defined using this equation. It is clear that this tensor would not represent the source term of the field equations but only the total stress energy of the interaction connection and matter frame fields.

It has been shown that the even part of the equations represent the classical interactions of gravitation and electromagnetism. We separate the equations with respect to the even subalgebra or subgroup, because this part represents the classical fields. The inclusion map allows the possibility of defining the pullback connection $i^*\omega$ and its composition with the homomorphisms *h* provides an so(3,1) valued connection which is compatible with a metric in space time. In other words, we have the $sl_1(2,C)$ forms,

$$i*\omega = AiI + \Gamma^a E_a \quad , \qquad (1.2)$$

$$i*\Omega = FiI + R^a E_a \quad . \qquad (1.3)$$

The curvature form *R* of the *Γ* connection corresponds to the Riemann curvature in standard spinor formulation. They obey the equations,

$$D*R = k\,{}^*J_+ \quad , \qquad (1.4)$$

$$DR = 0 \quad , \qquad (1.5)$$

which are not Einstein's equations but represent a spinor gravitation formulation equivalent to Yang's [4] theory restricted to its SO(3,1) subgroup. The restriction to the even part only gives vacuum solutions

Yang's gravitational theory may be seen as a theory of a connection in the principal bundle of linear frames $TM_m$ with structure group GL(4,R). In Yang's theory the group is taken to act on the tangent spaces to *M* and therefore is different from the theory under discussion. The connection in Yang's theory is not necessarily compatible with a metric on the base space *M* which leads to the known difficulties pointed out in the cited papers. Nevertheless when Yang's theory is restricted to its SO(3,1) subgroup its non metricity problems are eliminated. The well known homomorphism between this group and our even subgroup SL(2.C) establishes a relation between these two restrictions of the theories.

When we consider the external problem, that is space-time regions where there is no matter, and consider only the gravitational part, the equations obtained are similar to those of Yang's gravitational theory. All vacuum solutions of Einstein's equations are solutions of this equation for *J=0*. In particular the Schwarzchild solution is a solution to these equations and, therefore, the Newtonian motion under a *1/r* gravitational potential is obtained as a limit of the geodesic motion under the proposed equations. It should be pointed out the existence of a full Newtonian limit is not obvious.[5, 6]. There are additional vacuum solutions for these theories which are not solution for Einstein's theory. Fairchild [7] has shown that, for Yang's theory, [8] the eq. (0.1) is sufficient to rule out the spurious vacuum solutions found by Pavelle [9,10] and Thompson [11].

Nevertheless the interior problem provides a situation where there are essential differences between the unified physical geometry an general relativity. Also for the internal problem where the source $J$ is non-zero, the theory is essentially different from Yang's theory. Since in the physical geometry, the metric and connection remain compatible, the base space remains pseudo-Riemannian with torsion avoiding the difficulties discussed by Fairchild and others for Yang's theory..

The presence of a matter current term in the equation for the stress energy may afford the possibility of getting a geometric stress energy tensor that could enter in the equation for the metric, thus resembling Einstein's theory. This is the purpose of this paper.

## An Equation for the Einstein Tensor.

The equation

$$\mathrm{tr}\left[2\Omega_{\hat{\rho}\nu}\Omega^{\mu\nu} - \tfrac{1}{2}u_{\hat{\rho}}^{\mu}\Omega^{\kappa\lambda}\Omega_{\kappa\lambda}\right] = 2k\,\mathrm{tr}\left[e^{-1}\iota \circ u^{\mu}\nabla_{\hat{\rho}}e\right] \qquad (2.1)$$

defines a tensor field on $M$, a section in the tensor bundle over $M$. It is clear that the trace in the equation introduces a scalar product, the Killing metric $^{K}g$, in the fiber of the bundles and the result is valued in the tensor bundle. This Killing scalar product allows us to write the left hand side of this equation in terms of a summation over all components along the 15 generators of a base in the Lie algebra. We shall separate away the terms due to the sl(2,C) subalgebra, as follows,

$$\mathrm{tr}\left[2\Omega_{\rho\nu}\Omega_{\mu}{}^{\nu} - \frac{1}{2}g_{\rho\mu}\Omega^{\kappa\lambda}\Omega_{\kappa\lambda}\right] = {}^{K}g_{ab}\left[2\Omega^{a}{}_{\rho\nu}\Omega^{b}{}_{\mu}{}^{\nu} - \frac{1}{2}g_{\rho\mu}\Omega^{a\kappa\lambda}\Omega^{b}{}_{\kappa\lambda}\right] +$$
$$+ {}^{K}g_{\tilde{a}\tilde{b}}\left[2\Omega^{\tilde{a}}{}_{\rho\nu}\Omega^{\tilde{b}}{}_{\mu}{}^{\nu} - \frac{1}{2}g_{\rho\mu}\Omega^{\tilde{a}\kappa\lambda}\Omega^{\tilde{b}}{}_{\kappa\lambda}\right] \;, \qquad (2.2)$$

where the summation over Latin indices is restricted to the six components of the sl(2,C) subalgebra and the summation over Latin indices with tilde is over the nine components of the coset algebra. The latter terms, coming from the odd generators and the $\kappa_{5}$ generator, correspond to the stress energy tensor of the additional non gravitational interaction "coset" fields present in the theory, including the standard electromagnetic field. They all have the familiar quadratic structure in terms of the curvature components,

$$^{c}\Theta_{\rho\mu} \equiv \frac{-1}{4\pi}\,{}^{K}g_{\tilde{a}\tilde{b}}\left[\Omega^{\tilde{a}}{}_{\rho\nu}\Omega^{\tilde{b}}{}_{\mu}{}^{\nu} - \frac{1}{4}g_{\rho\mu}\Omega^{\tilde{a}\kappa\lambda}\Omega^{\tilde{b}}{}_{\kappa\lambda}\right] \;, \qquad (2.3)$$

and define a coset field stress energy tensor $^{c}\Theta$. Since the electromagnetic generator is compact, the Killing metric introduces a -1 and we must define $^{c}\Theta$ as shown so that the standard electromagnetic energy is positive definite.

The right hand side of the main equation should be interpreted as a stress energy tensor related to the matter current source. We define it as

$$^{j}\Theta_{\hat{\rho}}^{\mu} \equiv \alpha\,\mathrm{tr}\left[e^{-1}\iota \circ u^{\mu}\nabla_{\hat{\rho}}e\right] \;. \qquad (2.4)$$

Thus, we may write equation (2.12) in this manner

$$^K g_{ab} \left[ 2\Omega^a{}_{\rho\nu} \Omega^b{}_\mu{}^\nu - \frac{1}{2} g_{\rho\mu} \Omega^{a\kappa\lambda} \Omega^b{}_{\kappa\lambda} \right] = 8\pi \left( {}^j\Theta_{\rho\mu} + {}^c\Theta_{\rho\mu} \right) \quad . \tag{2.5}$$

The sl(2,C) generators corresponding to the left side in the last equation, the gravitational part, may be expressed in terms of the homomorphic so(3,1) generators $X_a$ (Lorentz rotation generators) as 4×4 matrices acting on the tangent bundle *TM*,

$$\begin{aligned} {}^K g_{ab} \left[ 2\Omega^a{}_{\rho\nu} \Omega^b{}_\mu{}^\nu - \frac{1}{2} g_{\rho\mu} \Omega^{a\kappa\lambda} \Omega^b{}_{\kappa\lambda} \right] &= \\ &= 2\, \mathrm{tr}\left[ X_a X_b \left( 2\Omega^a{}_{\rho\nu} \Omega^b{}_\mu{}^\nu - \frac{1}{2} g_{\rho\mu} \Omega^{a\kappa\lambda} \Omega^b{}_{\kappa\lambda} \right) \right] \\ &= 4 \left( \Omega^{\hat{\alpha}}{}_{\hat{\beta}\rho\nu} \Omega^{\hat{\beta}}{}_{\hat{\alpha}\mu}{}^\nu - \frac{1}{4} g_{\rho\mu} \Omega^{\hat{\alpha}}{}_{\hat{\beta}}{}^{\kappa\lambda} \Omega^{\hat{\alpha}}{}_{\hat{\beta}\kappa\lambda} \right) \end{aligned} . \tag{2.6}$$

The connection associated to the SO(3,1) group admits the possibility of torsion. We can further split away the torsion $\Sigma$ from the Levi-Civita connection,

$$\Gamma^\alpha_{\beta\mu} = \left\{ {\alpha \atop \beta\mu} \right\} + \Sigma^\alpha_{\beta\mu} \tag{2.7}$$

and express the curvature in terms if the Riemann tensor $R^\alpha{}_{\beta\mu\nu}$ and an explicit dependence on the torsion,

$$\Omega^\alpha{}_{\beta\kappa\lambda} = R^\alpha{}_{\beta\kappa\lambda} + Z^\alpha{}_{\beta\kappa\lambda} \quad, \tag{2.8}$$

where

$$Z^\alpha{}_{\beta\kappa\lambda} = \nabla_\kappa \Sigma^\alpha_{\beta\lambda} - \nabla_\lambda \Sigma^\alpha_{\beta\kappa} + \Sigma^\alpha_{\gamma\kappa} \Sigma^\gamma_{\beta\lambda} - \Sigma^\alpha_{\gamma\lambda} \Sigma^\gamma_{\beta\kappa} \quad . \tag{2.9}$$

Substitution in equation (2.5) gives an expression in terms of the Riemann tensor of the symmetric metric connection,

$$\begin{aligned} \left( 4\Omega^{\hat{\alpha}}{}_{\hat{\beta}\rho\nu} \Omega^{\hat{\beta}}{}_{\hat{\alpha}\mu}{}^\nu - g_{\rho\mu} \Omega^{\hat{\alpha}}{}_{\hat{\beta}}{}^{\kappa\lambda} \Omega^{\hat{\alpha}}{}_{\hat{\beta}\kappa\lambda} \right) &= \left( 4 R^{\hat{\alpha}}{}_{\hat{\beta}\rho\nu} R^{\hat{\beta}}{}_{\hat{\alpha}\mu}{}^\nu - g_{\rho\mu} R^{\hat{\alpha}}{}_{\hat{\beta}}{}^{\kappa\lambda} R^{\hat{\alpha}}{}_{\hat{\beta}\kappa\lambda} \right) + \\ & 4 Z^{\hat{\alpha}}{}_{\hat{\beta}\rho\nu} Z^{\hat{\beta}}{}_{\hat{\alpha}\mu}{}^\nu - g_{\rho\mu} Z^{\hat{\alpha}}{}_{\hat{\beta}}{}^{\kappa\lambda} Z^{\hat{\alpha}}{}_{\hat{\beta}\kappa\lambda} + 4 Z^{\hat{\alpha}}{}_{\hat{\beta}\rho\nu} R^{\hat{\beta}}{}_{\hat{\alpha}\mu}{}^\nu - g_{\rho\mu} Z^{\hat{\alpha}}{}_{\hat{\beta}}{}^{\kappa\lambda} R^{\hat{\alpha}}{}_{\hat{\beta}\kappa\lambda} + \\ & 4 R^{\hat{\alpha}}{}_{\hat{\beta}\rho\nu} Z^{\hat{\beta}}{}_{\hat{\alpha}\mu}{}^\nu - g_{\rho\mu} R^{\hat{\alpha}}{}_{\hat{\beta}}{}^{\kappa\lambda} Z^{\hat{\alpha}}{}_{\hat{\beta}\kappa\lambda} \end{aligned} . \tag{2.10}$$

The term in parenthesis in the right hand side is the expression previously developed by Stephenson [12] within the Yang [13] theory of gravitation,

$$H_{\rho\mu} = R^{\hat{\alpha}}{}_{\hat{\beta}\rho\nu} R^{\hat{\beta}}{}_{\hat{\alpha}\mu}{}^\nu - \frac{1}{4} g_{\rho\mu} R^{\hat{\alpha}}{}_{\hat{\beta}}{}^{\kappa\lambda} R^{\hat{\alpha}}{}_{\hat{\beta}\kappa\lambda} \tag{2.11}$$

and we may define a stress energy tensor associated to the torsion

$${}^t\Theta_{\rho\mu} \equiv \frac{-1}{2\pi}\left( \begin{array}{c} Z^{\hat{\alpha}}{}_{\hat{\beta}\rho\nu}Z^{\hat{\beta}}{}_{\hat{\alpha}\mu}{}^\nu - \frac{1}{4}g_{\rho\mu}Z^{\hat{\alpha}}{}_{\hat{\beta}}{}^{\kappa\lambda}Z^{\hat{\alpha}}{}_{\hat{\beta}\kappa\lambda} + Z^{\hat{\alpha}}{}_{\hat{\beta}\rho\nu}R^{\hat{\beta}}{}_{\hat{\alpha}\mu}{}^\nu - \frac{1}{4}g_{\rho\mu}Z^{\hat{\alpha}}{}_{\hat{\beta}}{}^{\kappa\lambda}R^{\hat{\alpha}}{}_{\hat{\beta}\kappa\lambda} \\ + R^{\hat{\alpha}}{}_{\hat{\beta}\rho\nu}Z^{\hat{\beta}}{}_{\hat{\alpha}\mu}{}^\nu - \frac{1}{4}g_{\rho\mu}R^{\hat{\alpha}}{}_{\hat{\beta}}{}^{\kappa\lambda}Z^{\hat{\alpha}}{}_{\hat{\beta}\kappa\lambda} \end{array}\right). \qquad (2.12)$$

In this manner we may write equation (2.12) as

$$H_{\mu\kappa} = 8\pi\left( {}^j\Theta_{\rho\mu} + {}^c\Theta_{\rho\mu} + {}^t\Theta_{\rho\mu} \right). \qquad (2.13)$$

There is an alternate expression for $H$, given by Fairchild [14], obtained by decomposing the Riemann tensor in terms of the Weyl tensor, the Ricci tensor and the scalar $R$.

$$H_{\rho\mu} = \frac{5R}{3}\left(R_{\rho\mu} - \frac{1}{4}g_{\rho\mu}R\right) - 2R_{\rho\sigma}R_\mu^\sigma + \frac{1}{2}g_{\rho\mu}R_{\kappa\lambda}R^{\kappa\lambda} + C^\kappa{}_{\mu\rho\lambda}R^\lambda_\kappa, \qquad (2.14)$$

which implies that it may be written in terms of the Einstein tensor $G_{\mu\nu}$ as

$$H_{\rho\mu} = \frac{5R}{3}\left(G_{\rho\mu} + \frac{1}{4}g_{\rho\mu}R\right) - 2R_{\rho\sigma}R_\mu^\sigma + \frac{1}{2}g_{\rho\mu}R_{\kappa\lambda}R^{\kappa\lambda} + C^\kappa{}_{\mu\rho\lambda}R^\lambda_\kappa. \qquad (2.15)$$

Because of the nonlinearity of the gravitational equations there is a contribution to the source of gravitation from the metric field itself. We may define a stress energy tensor associated to the metric $g$ (gravitational field),

$${}^g\Theta_{\mu\nu} \equiv \frac{1}{8\pi}\left[ 2R_{\rho\sigma}R_\mu^\sigma - g_{\rho\mu}\left(\frac{5R^2}{12} + \frac{R_{\kappa\lambda}R^{\kappa\lambda}}{2}\right) - C^\kappa{}_{\mu\rho\lambda}R^\lambda_\kappa \right] \qquad (2.16)$$

The different $\Theta$ terms are grouped together, defining a total generalized geometric stress energy tensor designated by $\Theta_{\mu\nu}$ to distinguish it from the standard $T_{\mu\nu}$. This tensor includes contributions from the matter current and the total field energy. The equation (2.12) may be written

$$\frac{5R}{3}G_{\mu\nu} = 8\pi\Theta_{\mu\nu} = 8\pi\left( {}^g\Theta_{\mu\nu} + {}^t\Theta_{\mu\nu} + {}^c\Theta_{\mu\nu} + {}^j\Theta_{\mu\nu} \right). \qquad (2.17)$$

We have a generalized Einstein equation with a geometric stress energy tensor.

If there is zero matter current, zero coset fields *and* zero torsion, there is only ${}^g\Theta$ and, we obtain the Stephenson-Yang equation

$$\frac{5R}{3}G_{\mu\nu} - 8\pi\,{}^g\Theta \equiv H_{\mu\nu} = 0. \qquad (2.18)$$

Fairchild has shown [11] that this equation implies that the only empty space solutions are the Einstein spaces, ruling out the exceptional static spherically symmetric solutions given by Thomson [15] and Pavelle [16]. If $\Theta$ is non zero, then $R$ should be non zero and we may write a generalized Einstein equation,

$$G_{\mu\nu} = R_{\mu\nu} - \frac{1}{2}g_{\mu\nu}R = 8\pi\frac{3}{5R}\Theta_{\mu\nu}. \qquad (2.19)$$

## Equations for a Geometric Internal Schwarzschild Solution.

If we assume spherical symmetry, the metric takes the form,

$$d\tau^2 = e^{2\Phi}dt^2 - e^{2\Lambda}dr^2 - r^2 d\Omega \;, \tag{3.1}$$

Under this condition, an internal solution for Einstein's theory may be determined by solving the following equations:

1. The time-component of the field equations $G_{00}$;
2. The radial component of the field equations $G_{11}$;
3. The conservation of the energy momentum tensor;
4. The equations of state of matter.

In our theory we have essentially the same requirements.

First we may calculate the necessary values of $G_{\mu\nu}$ with respect to an orthonormal frame [14, p. 602]. For the time component we have,

$$G_{\hat{o}\hat{o}} = \left(\frac{1}{r^2} - \frac{e^{-2\Lambda}}{r^2} - \frac{1}{r}\frac{de^{-2\Lambda}}{dr}\right) = \frac{1}{r^2}\frac{d}{dr}\left(r(1-e^{-2\Lambda})\right) = 8\pi\frac{3\Theta_{\hat{o}\hat{o}}}{5R} \;, \tag{3.2}$$

where the right hand side is only a function of $r$ and may be integrated,

$$\int_0^r dr\, 4\pi r^2 \left(\frac{3\Theta_{\hat{o}\hat{o}}}{5R}\right) = \int_0^r \frac{dr}{2}\frac{d}{dr}\left(r(1-e^{-2\Lambda})\right) = \frac{r}{2}(1-e^{-2\Lambda}) \equiv GM(r) \;, \tag{3.3}$$

defining a macroscopic mass $M$, using a constant $G$ that we keep arbitrary. The spherical potential inside the matter distribution may be taken as

$$\varphi = -\frac{GM(r)}{r}. \tag{3.4}$$

obtaining the following expression inside the matter distribution,,

$$e^{-2\Lambda} = 1 - \frac{2GM}{r} = 1 + 2\varphi \;. \tag{3.5}$$

In addition we have for the radial component,

$$G_{\hat{i}\hat{i}} = \left(\frac{e^{-2\Lambda}}{r^2} - \frac{1}{r^2} + \frac{2e^{-2\Lambda}}{r}\frac{d\Phi}{dr}\right) = 8\pi\frac{5\Theta_{\hat{i}\hat{i}}}{5R} \;, \tag{3.6}$$

which may be solved for the derivative of $\Phi$,

$$\frac{d\Phi}{dr} = \frac{GM + 4\pi r^3 \left(3\Theta_{\hat{\imath}\hat{\imath}}/5R\right)}{r(r-2GM)} \quad , \tag{3.7}$$

where we have use the previous definition of *M*.

In second place we have the conservation of the Einstein tensor, with respect to the induced Levi-Civita connection in the bundle *TM*, which implies the similar conservation of the right hand side of equation (2.18)

$$\nabla_\mu \left(\frac{\Theta^{\rho\mu}}{R}\right) = 0 \quad . \tag{3.8}$$

Finally, instead of equations of state, we have that the expression for the geometric *Θ* is determined by solutions of the three coupled nonlinear geometric equations, including the one that reduces to the quantum Dirac equation. Since all the equations may be derived from a variational principle, we expect that they are consistent with each other and together with appropriate boundary condition would determine a fully geometric internal Schwarzschild solution. It should be noted here that this is a consequence of having a geometric energy momentum as Einstein aspired for a fully geometric theory. Alternately, we may assume that although the source is fully geometric, it should have the phenomenological properties known for macroscopic matter, for example the known equations of state for a fluid, and in this phenomenological approximation, reduce the problem to the usual general relativistic macroscopic internal problem.

This internal spherical solution should be matched with the external Schwarzschild solution at the spherical boundary. Due to the vacuum equations outside the spherical matter distribution, it is known that, in the exterior

$$g_{00} = e^{2\Phi} = e^{-2\Lambda} = 1 + 2\varphi \quad , \tag{3.9}$$

which relates the external Schwarzschild metric and the spherical gravitational potential, where *GM* is constant related to the total mass inside the spherical boundary. Thus, the corresponding matching exterior solution satisfies the physical requirements for gravitational field of a massive spherical body.

## The Newtonian Limit.

If we make the usual assumptions $v/c \ll 1$, $\varphi \ll 1$ [17, p. 412], that lead to the Newtonian limit in Einstein's theory, the Newtonian limit is obtained from eq. ((2.18)). The non vanishing components $\Gamma_{00}{}^a$ of the limit connection give the non vanishing components of the Riemann tensor,

$$R_{00} = R^a_{0a0} = \partial_a \Gamma^a_{00} = \delta^{ab} \partial_a \partial_b \varphi \quad . \tag{4.1}$$

Also from the Einstein equation we get the equation for the only component,

$$R_{00} = \frac{1}{2}\left(8\pi \frac{3\Theta_{\hat{0}\hat{0}}}{5R}\right) \quad . \tag{4.2}$$

Together the two equations result in Poisson's equation,

$$\partial_a \partial^a \varphi = 4\pi \left(\frac{3\Theta_{\hat{0}\hat{0}}}{5R}\right) = 4\pi \rho G \quad . \tag{4.3}$$

where clearly the Newtonian limit of the expression in parenthesis must be interpreted as the Newtonian mass density $\rho$. If, $\Phi = \Lambda$, the density is in agreement with the definition of the total mass $M$ of the spherical solution given in the previous section.

## Conclusions.

The general relativity Einstein equations with cosmological constant follow from the stress energy equation for empty space. For non empty space we obtain a generalized Einstein equation, eq. (2.18) relating the Einstein tensor $G_{\mu\nu}$ to a geometric stress energy tensor $\Theta_{\mu\nu}$. For spherical symmetry the mass of a body may be defined in terms of energy-mass integrals, as in Einstein's theory. The matching exterior solution is in agreement with the standard relativity tests. In addition, there is a Newtonian limit where we obtain the corresponding Poisson's equation.